\documentclass[journal=jpclcd,manuscript=article,layout=onecolumn]{achemso}
\setkeys{acs}{articletitle = true}

\usepackage{graphicx}
\usepackage{amssymb}
\usepackage[toc,page]{appendix}
\usepackage{color}

\newcommand{\R}{\mathbf{r}}
\def\dd{{\rm d}^3}

\newcommand{\onlinecite}[1]{\citenum{#1}}

\title{Semilocal Pauli-Gaussian Kinetic Functionals for Orbital-Free Density Functional Theory Calculations of Solids }
\author{Lucian A. Constantin}
\affiliation{Center for Biomolecular Nanotechnologies @UNILE, Istituto Italiano di Tecnologia, Via Barsanti, I-73010 Arnesano, Italy}
\author{Eduardo Fabiano}
\affiliation{Institute for Microelectronics and Microsystems (CNR-IMM), Via Monteroni, Campus Unisalento, 73100 Lecce, Italy and Center for Biomolecular Nanotechnologies @UNILE, Istituto Italiano di Tecnologia, Via Barsanti, I-73010 Arnesano, Italy}
\author{Fabio Della Sala}
\email{fabio.dellasala@unisalento.it}
\affiliation{Institute for Microelectronics and Microsystems (CNR-IMM), Via Monteroni, Campus Unisalento, 73100 Lecce, Italy and Center for Biomolecular Nanotechnologies @UNILE, Istituto Italiano di Tecnologia, Via Barsanti, I-73010 Arnesano, Italy}

\date{\today}

\begin{document}

\begin{tocentry}
\includegraphics{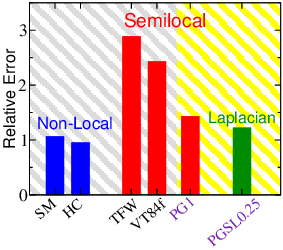}
\end{tocentry}

\begin{abstract}
Kinetic energy (KE) approximations are key elements in orbital-free density
functional theory. To date, the use of non-local functionals, possibly employing system
dependent parameters, has been considered mandatory in order to obtain
satisfactory accuracy for different solid-state systems, whereas semilocal
approximations are generally regarded as unfit to this aim.
Here, we show that instead properly constructed semilocal approximations, the Pauli-Gaussian (PG)
KE functionals, 
especially at the Laplacian-level of theory, can
indeed achieve similar accuracy as non-local functionals and can be 
accurate for both metals and semiconductors, without 
the need of system-dependent parameters.
\end{abstract}

%\pacs{71.10.Ca,71.15.Mb,71.45.Gm}

%\maketitle

The kinetic energy (KE) functional is a fundamental quantity 
in electronic structure
theory. It plays a prominent role in subsystem and embedding theories
\cite{elliott10,huang11,jacob14,wesochemrev15,sun16}, 
hydrodynamic models \cite{banerjee00,toscano15,ciraci16}, 
information theory \cite{trickey11,nagy14}, 
machine learning techniques for 
Fermionic systems \cite{snyder12,yao16,seino18}, 
potential functional theory \cite{cangi2011electronic}, various extensions of the Thomas-Fermi (TF) theory 
\cite{lehtomaki2017self,ribeiro2018deriving},  
and
especially orbital-free density functional theory (OF-DFT)
\cite{ofdft_book,wang2002orbital,lehtomaki14,karasiev2014progress,karasiev2015chapter,zavodinsky17,carter18}.
The applicability of these methods is strongly limited by 
the lack of accurate and simple KE approximations.  

The OF-DFT allows an efficient
description of the ground state
of any electronic system via the solution of the Euler equation \cite{levy1988exact,levy1984exact}
\begin{equation}
\frac{\delta T_s[n]}{\delta n} + v_{ext}(\R) + \int\frac{n(\R')}{|\R-\R'|}\dd\R' + \frac{\delta E_{xc}[n]}{\delta n} 
= \mu\ ,
\label{e1}
\end{equation}
where $n$ is the ground-state electron density,
$T_s$ is the non-interacting KE functional, $v_{ext}$ is the
external (e.g. nuclear) potential, $E_{xc}$ is the exchange-correlation
functional \cite{scuseriaREVIEW05,della2016kinetic}, and $\mu$ is the
chemical potential.

The vast majority of the OF-DFT calculations employ
non-local (or two-point) KE functionals,
that are rather accurate and display a logarithmic $\mathcal{O}(N\ln(N))$ 
scaling behavior with system size $N$. However, these functionals are based on 
the Lindhard response function 
of the non-interacting homogeneous electron gas (HEG)
\cite{wang1992kinetic,perrot1994hydrogen,smargiassi1994orbital,wang98,pavanello2018,constantin18}. 
Thus, 
%NEW
most of them
%they 
depend 
on the average density in the unit cell $n_0$ and therefore they are not adequate for finite systems or 
even for anisotropic solid-state systems such as interfaces, surfaces or layered 
materials, where $n_0$ may be not well defined \cite{xia2012can,xia12,shin14} or may be not representative 
for the system. Moreover, very accurate results are often obtained only using system-dependent parameters\cite{huang10,shin14,pavanello2018,constantin18}.

On the other hand, unlike for the exchange-correlation energy case where semilocal 
approximations \cite{perdewPRL96,della2016kinetic,scuseriaREVIEW05}
have experienced a huge success in the context of Kohn-Sham (KS) DFT \cite{burke2012perspective}, semilocal KE 
approximations,
%NEW 
which scale linearly with system size,
%which give a fast linear-scaling with system size 
and can be easily implemented in both real-space and wave-vector formalisms, are barely used in OF-DFT calculations.
This traces back to the fact that the actual 
state-of-the-art semilocal 
functionals (e.g. VT84f \cite{karasiev2013nonempirical}, vWGTF1/vWGTF2 \cite{xia2015single})
may encounter severe failures for various systems (e.g. semiconductors) and 
properties (e.g. bulk modulus, vacancy energy \cite{xia2015single}).

In this Letter, we show that this limitation is not a fundamental feature of the 
semilocal functionals but it is just related to the approximations employed so far.
Indeed, we show that even a simple combination of the local
Thomas-Fermi (TF) \cite{thomas1927calculation,fermi1927metodo} and
gradient-dependent von Weizs\"{a}cker (W) \cite{weizsacker1935theorie}
functionals can outperform the actual state-of-the-art semilocal functionals
in solid-state calculations. Moreover, we show that a   
a simple non-empirical Laplacian-level semilocal KE functional can easily 
% pair the accuracy of
%NEW
approach the accuracy of
non-local KE approximations, achieving a broad accuracy and applicability.
These results shed a completely new light on the topic of KE functionals, showing
that with a careful development semilocal KE functionals 
%NEW
can be  applied with good accuracy in large-scale OF-DFT applications.

To this purpose, we have performed OF-DFT 
calculations for simple metals \cite{xia2015single} (Li, Mg and Al, in simple-cubic (sc), 
face-centered-cubic (fcc), and body-centered-cubic (bcc) configurations)
and III-V semiconductors \cite{shin14,huang10} (AlP, AlAs, AlSb, GaP, GaAs,
GaSb, InP, InAs, and InSb with the cubic zincblende unit cell), 
comparing our results to KS-DFT values obtained using the
same computational set up.
%NEW
These systems have been largely employed to asses the accuracy of non-local KE functionals \cite{xia2015single,huang10,shin14,xia12,pavanello2018}.
We considered four properties: cell volume ($V_0$), 
bulk modulus ($B$), total energy at equilibrium volume ($E_0$) 
and density error ($D_0$).
The first three properties have been previously considered in the assessment of
functionals \cite{xia2015single,huang10,shin14,xia12,pavanello2018}. The last quantity is defined as
\begin{equation}
D_0=\frac{1}{N_e} \int |n^{KS}({\bf r})-n^{OFDFT}({\bf r})| \dd\R,
\label{e3}
\end{equation}
and it is computed at the KS lattice constant for both KS and OF-DFT. Here $N_e$ is the electron number in the unit cell.
The density error is a very hard test for the quality of the KE functional, describing how well the OF-DFT 
calculations converge to the exact density.
For each quantity $p \in \{V_0,\ B,\ E_0,\ D_0\}$, we considered the mean absolute relative error (MARE$_p$) with respect to the 
reference KS values averaging over all systems (metals or semiconductors).
Finally, in order to have a global indicator for all the properties, we
considered a relative MARE (RMARE) obtained normalizing 
to the average values of the Smargiassi and Madden (SM) \cite{smargiassi1994orbital}
and the Huang and Carter (HC) \cite{huang10} functionals, i.e.
\begin{equation}
RMARE = \sum_p \frac{MARE_p}{(1/2)( MARE_p^{SM}+MARE_p^{HC})}\ .
\label{e4}
\end{equation}
The HC (with universal parameters $\lambda=0.01177$ and $\beta=0.7143$
\cite{shin14} ) and SM functionals have been chosen as references, since
they bind both metals and semiconductors \cite{constantin18} 
and do not employ system-dependent parameters 
%NEW
(actually, HC does not even depend on the average density $n_0$).
% and do not employ system-dependent parameters others than the average density $n_0$.
Thus, a functional with RMARE=1 shows a performance between HC and SM.

\begin{figure}[tb]
\includegraphics[width=\columnwidth]{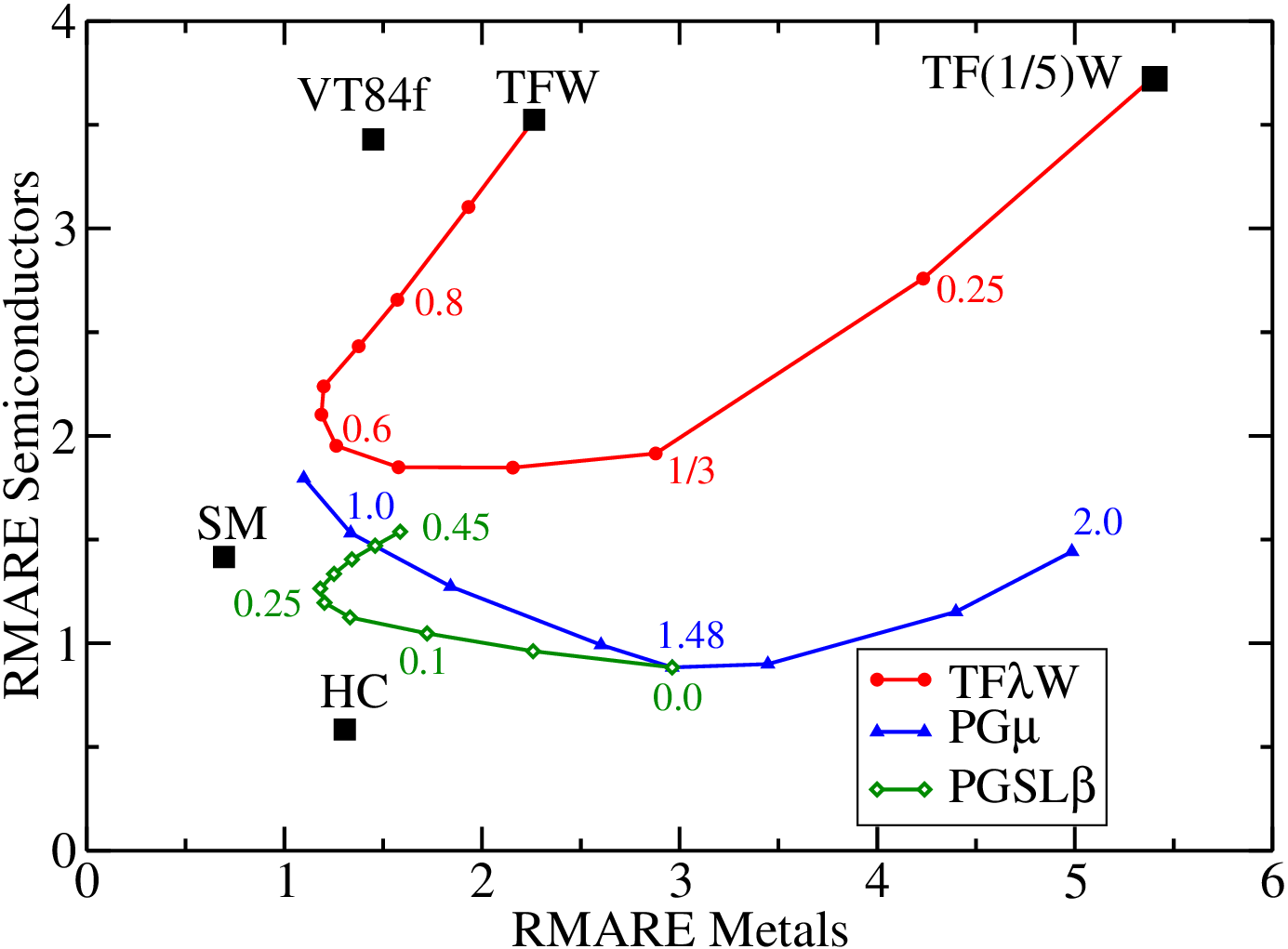}
\caption{\label{fig_main1} RMARE of metals vs RMARE of semiconductors for different functionals.}
\end{figure}

Figure \ref{fig_main1} reports the RMAREs for metals and semiconductors
of different semilocal functionals commonly used in OF-DFT calculations, namely VT84f \cite{karasiev2013nonempirical}, 
TFW, TF(1/5)W (the latter being combinations of the TF functional with
W and W/5, respectively), 
as well as for the non-local references HC and SM.
Note that other functionals from literature, not only
semilocal, but also WT \cite{wang1992kinetic}, WGC \cite{wang1999orbital}, and
vWGTF \cite{xia2015single} do not bind
semiconductors \cite{constantin18,shin14}, so they cannot be included in Fig. \ref{fig_main1}.
Clearly the non-local functionals give the lowest RMAREs, with 
SM being especially accurate for metals and HC
for semiconductors.
On contrast, the semilocal functionals (TF(1/5)W, TFW, VT84f) perform much
worse (VT84f is quite accurate only for metals).

Next we consider more general semilocal functionals.
Any Generalized Gradient Approximation (GGA)
or Laplacian-level meta-GGA semilocal KE functional that
behaves correctly under the uniform density scaling
(i.e. $T_s[n_\gamma(\R)]=\gamma^2T_s[n(\R)]$ \cite{levy1985hellmann,fabianoPRA13}, where 
$n_\gamma(\R)=\gamma^3n(\gamma \R)$, with $\gamma\ge 0$)
can be written
\begin{equation}
T_s[n] = T_s^{W}+\int\tau^{TF} F_s^p(s,q) \dd\R,
\label{e2}
\end{equation}
where $\tau^{TF}=(3/10)k_F^2 n$ is the TF KE density
\cite{thomas1927calculation,fermi1927metodo}, with 
$k_F=(3\pi^2n)^{1/3}$ being the Fermi wave vector,
$s=|\nabla n|/(2k_Fn)$ and $q=\nabla^2 n/(4k_F^2n)$ are the reduced gradient and Laplacian,
$T_s^{W}=\int\tau^{TF}(5s^2/3)\dd\R$ is the von Weizs\"{a}cker
\cite{weizsacker1935theorie}
kinetic energy (which is exact for one and two electron systems, as well as for any bosonic system),
and  $F_p$ is the Pauli KE enhancement factor \cite{levy1988exact}.
The exact condition $T_s\ge T_s^W$ requires that $F_p>0$.
However, this constraint is satisfied only by
few semilocal approximations \cite{xia2015single,karasiev2013nonempirical,perdew2007laplacian},
and it is even violated  by some non-local KE functionals \cite{blanc2005nonlinear}.

{\bf The rehabilitation of semilocal KE functionals for OF-DFT.}
In a first attempt, we have considered the
family of functionals TF$\lambda$W which are defined by
\begin{equation}
F_{s,\lambda}^p = 1 + (\lambda-1)\frac{5}{3}s^2\ .
\end{equation}
This class of functionals has been 
investigated for atomic/molecular systems \cite{chan2001thomas,leal2015}
but not for bulk systems with pseudopotentials.
The performance for various values of $\lambda$ is reported in 
Fig. \ref{fig_main1}.
The first interesting result of this Letter, is that  
the RMAREs can be strongly reduced varying $\lambda$, reaching
for $\lambda=0.6$ a RMARE$\approx$1.2, and 1.9 for metals and semiconductors, respectively.
Thus a very simple functional, TF(0.6)W is already better than the
current semilocal state-of-the-art (VT84f).

\begin{figure}[tb]
\includegraphics[width=\columnwidth]{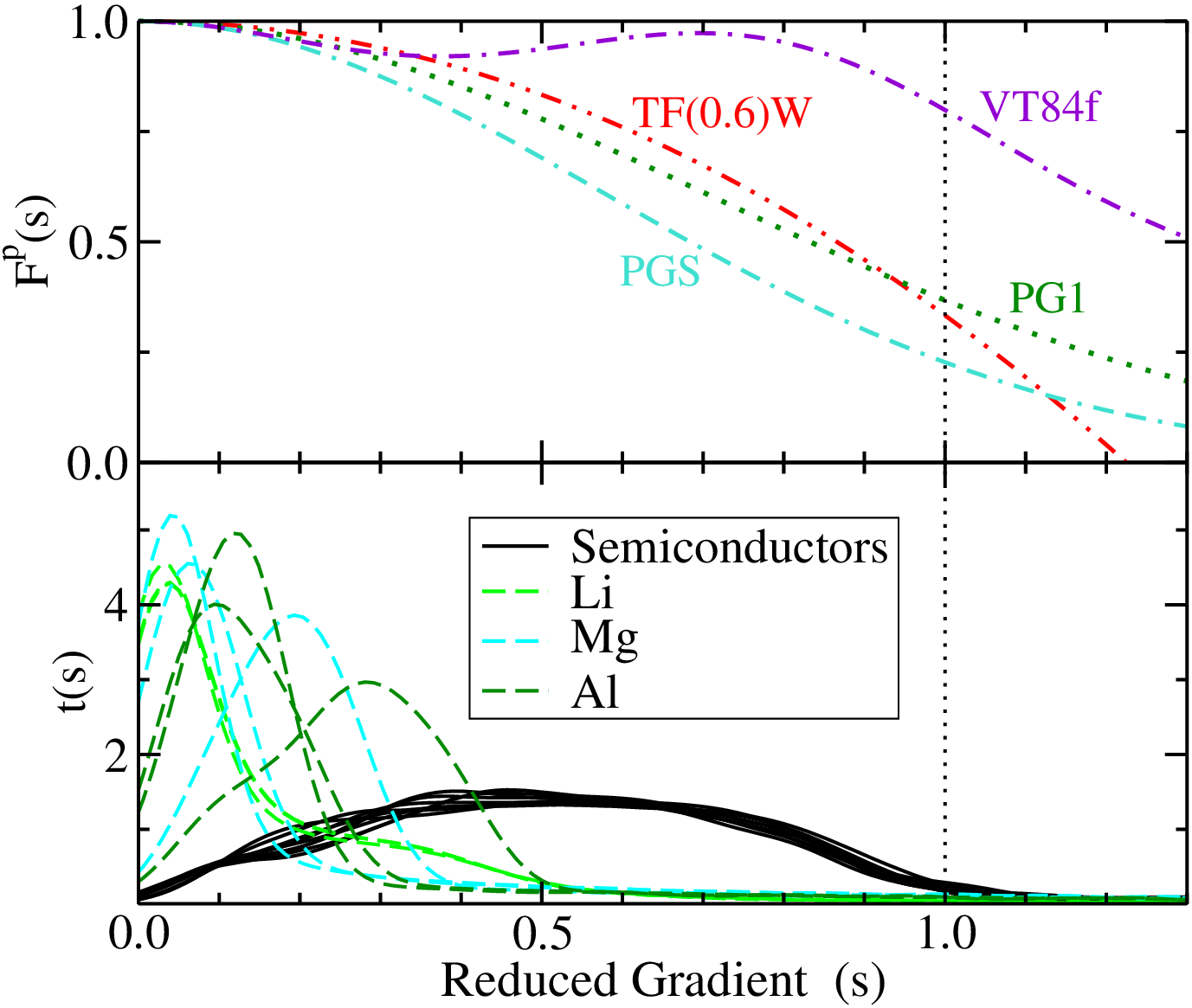}
\caption{\label{fig_sd} Upper panel: Pauli enhancement factor for different functionals;
Lower panel: s-decomposition of the TF kinetic-energy for different systems.}
\end{figure}

However, the Pauli enhancement factor of TF$\lambda$W becomes negative 
whenever $s^2\ge0.6/(1-\lambda)$.
The relevant values of $s^2$ for solids are comprised in the
interval [0:1], as shown in Fig. \ref{fig_sd} by 
an $s$-decomposition of the TF KE energy $t[n](s)$, so that\cite{laricchia2011}
\begin{equation}
T_s[n]=\int_0^{+\infty}ds\; t[n](s)\; F_s(s)\ .
\label{e6}
\end{equation}

Thus, as shown in Fig. \ref{fig_sd}, the Pauli enhancement factor for TF(0.6)W is always positive, which
could explain its relative good 
%NEW
 performance. 
Nevertheless, the tendency of $F_{s,\lambda}^p$ to become rather small and with high slope at high $s$
may limit its performance in particular for semiconductors, that are characterized
by larger values of $s$.

To overcome this problem, a new class of functionals (named Pauli-Gaussian, PG$\mu$) can be 
constructed considering the positive-defined Pauli enhancement factor
\begin{equation}
\label{eq:pg}
F_{s}^p(s)=e^{-\mu s^2} \ge 0 
\;.
\end{equation}
When $\mu\approx 1.96 \lambda^2 - 5.33 \lambda +3.37$, then $F_{s,PG\mu}^p\approx F_{s,\lambda}^p$ 
for $0\le s\le 1$. 
As expected, the PG$\mu$ functionals are significantly better than 
TF$\lambda$W for semiconductors and have similar accuracy for simple metals, 
as shown in Fig. \ref{fig_main1}. 
The best global performance is obtained approximately with $\mu=1$ (which defines the
PG1 functional).
However, this functional does not satisfy the second-order gradient expansion (GE2), which
is instead an important exact property to be retained \cite{Kirz57}.
On the other hand, the functional with
%NEW
 $\mu=40/27\approx 1.48$ 
 (named PGS, 
%NEW
 from Pauli-Gaussian Second order) satisfies
the GE2 constraint, is very good for semiconductors
but quite bad for metals (see Fig.  \ref{fig_main1}).

While we do not exclude that further optimizations of $F_s^p(s)$
could lead to improved results, here we follow a different path,
and we move to the Laplacian-meta-GGA level of theory, 
considering the PGSL$\beta$ 
%NEW
 (from Pauli-Gaussian Second order and Laplacian) class of 
functionals, defined by
\begin{equation}
\label{pgsl}
F_s^p(s,q)=e^{-40/27 s^2}+F(q) \approx e^{-40/27 s^2}+\beta q^2+ ...
\end{equation}
In Eq. (\ref{pgsl}), $F(q)$ represents a generic function of the reduced Laplacian, which can be
expanded in Taylor series because $q$ is always small in solids.
The linear term in $q$ does not contribute to the energy nor to the potential 
\cite{della2016kinetic}, thus the $q^2$ term considered here is just the
lowest-order correction.
Note that the $q^2$ term in Eq. (\ref{pgsl}) would cause a divergence near
the cusp of an electronic density (i.e. at the nuclei). However,
this shortcoming is not present using pseudopotentials.
In the tail of an exponentially decaying density (e.g. far away from an atom or a surface)
$q$ is diverging, but the full kinetic contribution
$\tau^{TF}q^2$ is still exponentially decaying, being integrable \cite{laricchia2013laplacian,engel1991theory}.
%NEW
We also recall that the HEG fourth-order gradient expansion has been successfully applied to
metallic clusters in the OF-DFT context \cite{engel1991theory}.

%The $q^2$-term is the only fourth-order term that accounts for the HEG linear response \cite{constantin2017jellium}.
%The next sixth-order linear response term is of the form $\sim |\nabla \nabla^2 n|^2$, and it is not integrable for
%finite systems \cite{yan1997numerical}.
%}

To fix the $\beta$ coefficient in a non-empirical way we could note that setting
$\beta=8/81\approx 0.1$ in Eq. (\ref{pgsl}), the corresponding functional recovers the fourth-order linear response of the non-interacting HEG, i.e. 
the Lind4 functional, see Ref. \cite{constantin2017jellium}. 
%
%5We presented this functional in Ref. \onlinecite{sof}.

As shown in Fig. \ref{fig_main1}, 
%we see that also for the structural and energy properties
the Laplacian plays an important role.
Varying $\beta$ from 0 to 0.25 we observe a large improvement of
simple metal properties (the RMARE decreases from 3 to 1.2) and correspondingly
a very small change in the accuracy of semiconductors (RMARE increases from 0.9 to 1.2).
The functional PGSL0.25 ($\beta=0.25$) 
is competitive with HC and SM, being only 20\% worse than their average.
This is the second main result of this work: a very simple Pauli KE functional 
(a sum of an exponential and a Laplacian-dependent quadratic term)
is almost as accurate as complicated and sophisticated non-local expressions. 
% This fact implies the rehabilitation of semilocal KE functionals for OF-DFT. 

The good 
%NEW
 performance
 of the  PGSL0.25 functional can be rationalized      
%Anyway, an even better guess for $\beta$ can be obtained requiring an accurate HEG linear response at 
%any wave vector, following the same spirit used for two-point KE functionals, and    
considering the exact density-density response function the HEG 
system which is $\chi(\eta)=(k_F/\pi^2)(1/F^{Lind}(\eta))$, with $\eta=k/2k_F$ and $F^{Lind}$ being the 
dimensionless wave vector and the Lindhard function \cite{lindhard1954properties}, respectively.
\begin{figure}[tb]
\includegraphics[width=\columnwidth]{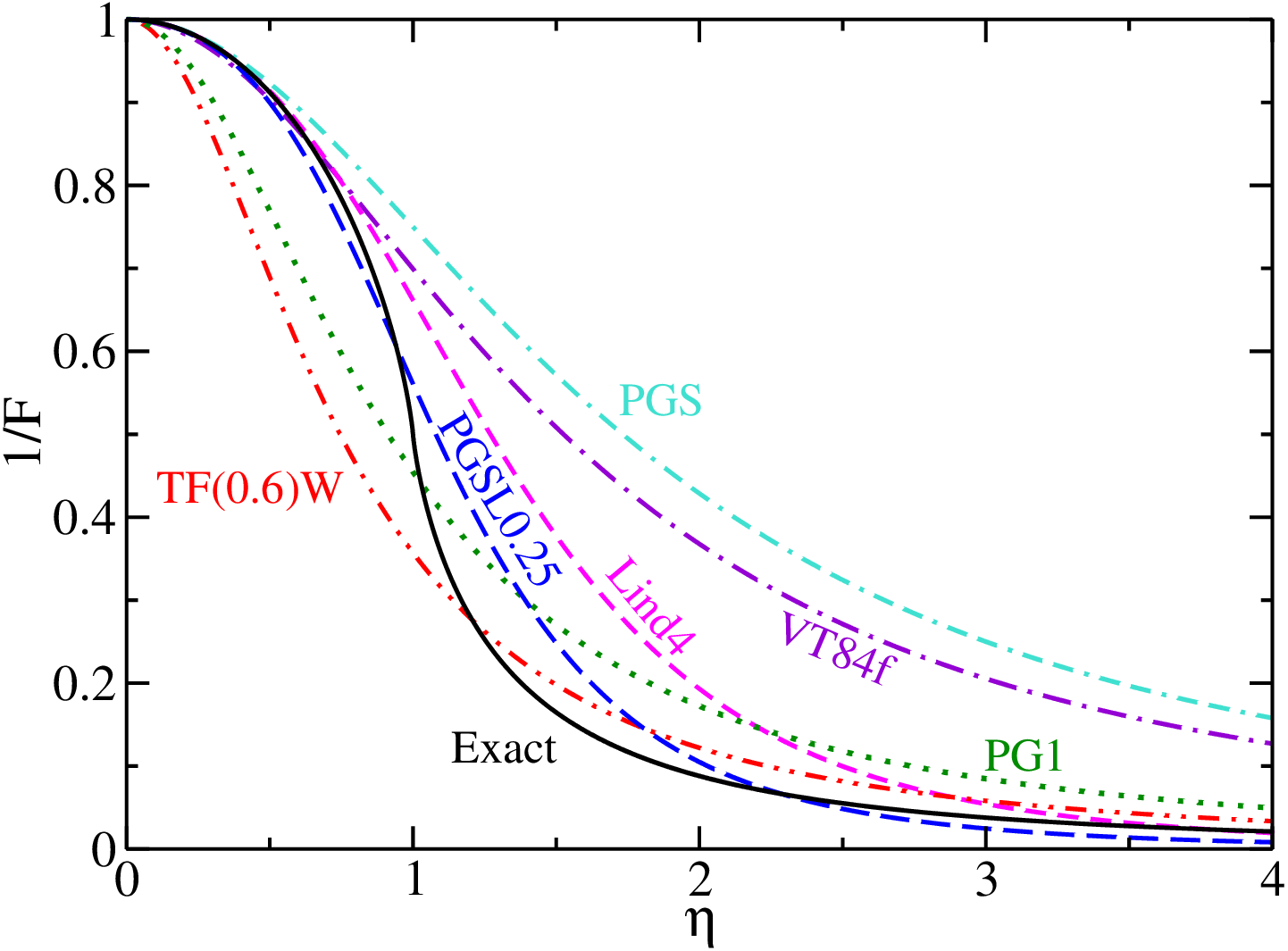}
\caption{\label{f3} Linear response function of non-interacting HEG ($\pi^2 \chi(\eta)/k_F$),
computed from various KE functionals.}
\end{figure}
Figure \ref{f3} reports the behavior of the linear response function for different functionals.
We see that TF(0.6)W and PG1 are good for $\eta>1$, while PGS and VT84f are good for  $\eta<1$: thus none 
of the gradient-dependent functionals considered is able to reproduce the exact behavior for all $\eta$.
On the other hand, a significant improvement is obtained by moving to the Laplacian level of theory: the PGSL0.25 
functional describes accurately both the low- and the high-$\eta$ regions.
% and provides a realistic description of the overall behavior.

%
\begin{table*}[tb]
\caption{\label{tab_Results} Mean absolute relative errors (in \%) for equilibrium volumes ($V_0$), bulk 
moduli ($B$), total energies ($E_0$) and density errors ($D_0$) 
of the simple metals and semiconductors test sets. 
%The full results can be found in the supporting information (Ref. \onlinecite{suppinf}.)
The best results from semilocal functionals are shown in bold style, and 
a star indicates
accuracy comparable or better than  $1/2$(HC+SM).
}
%\begin{ruledtabular}
\begin{tabular}{lrrrrrrrr}
\hline\hline
 & \multicolumn{4}{c}{Simple Metals} & \multicolumn{4}{c}{Semiconductors} \\
\cline{2-5}\cline{6-9  }
                  & $V_0$ & $B$           & $E_0$&  $D_0$   &  $V_0$         & $B$          & $E_0$         & $D_0$ \\
%  TF(1/5)W &        7.8 &       23.7 &       3.23 &       12.0 &       32.5 &       68.7 &       1.65 &       15.5 \\ 
  TFW     &        4.8 &       20.3 &       0.91 &        4.4 &        5.7 &       30.8 &       5.35 &       23.4 \\ 
  VT84f     &      *4.4 &       15.8 & *{\bf 0.14} &       3.9 &       10.5 &       63.5 &       3.56 &       22.1 \\ 
  TF(0.6)W &  *{\bf 2.6}&       12.1 &     *0.35 &        3.1 &         *2.9 &   *9.2 &       2.99 &       16.1 \\ 
  PG1      &      *3.6 &  *{\bf 7.3} &      0.45 &        3.7 &        *3.1 &   *8.2 &       2.03 &       14.7 \\ 
%  PGSL0.1  &       5.8 &        8.9  &      0.45 &        5.1 &        3.4  &    10.1  &     0.88 &       12.6 \\ 
  PGSL0.25 &      *4.4 &  *{\bf 7.4} &    *0.35 &    \bf{2.8} & *{\bf  2.5} &   *{\bf 5.7} & {\bf 1.62} & {\bf 13.3} \\ 
% 1.48_0.00 &       7.2 &       10.6 &       1.39 &        7.8 &        4.3 &       13.3 &       0.21 &       12.9 \\ 
\hline
$1/2$ (HC+SM)   &        4.3 &        7.8 &       0.34 &        1.7 &        4.5 &       21.1 &       0.63 &        7.9 \\ 
\hline
   SM        &        3.4 &       *4.2 &       0.21 &       *1.4 &        7.6 &       37.7 &       0.80 &       *7.6 \\ 
    HC        &        5.0 &       11.5 &       0.47 &        2.0 &       *1.6 &       *4.5 &      *0.46 &        8.3 \\ 
\hline\hline
\end{tabular}
%\end{ruledtabular}
\end{table*}

In more details, in Table \ref{tab_Results} we report the accuracy of the various functionals for the 
all the considered properties, separately (results for all systems are reported in Supporting Information). 
For simple metals all the semilocal functionals introduced in this Letter give quite accurate results,
being comparable to the non-local KE functionals.
%(with some exceptions for bulk moduli and densities).
On the other hand, semiconductors are more difficult systems and all the MARE are larger (but the errors for $V_0$, which are comparable).
Among the semilocal functionals considered, only PGSL0.25 gives always consistently accurate results.
In particular, it performs better than 1/2(SM+HC) for both lattice constants
and bulk moduli; for energies and densities (the hardest test), it is twice worse than
1/2(HC+SM), but still much more accurate than any other semilocal functionals.
(A comparison of GaAs densities is reported in the Supporting Information.)

Finally, in order to verify the broader applicability of
the PGSL0.25 functional, we considered additional systems and properties.
In Table \ref{tab:siv} we report the MARE for 
several silicon phases (sc, fcc, bcc and cd, for cubic diamond) and vacancy formation energies for 
fcc Al, hcp Mg, and bcc Li \cite{xia2015single}.
Again, for all properties, 
PGSL0.25 is the best semilocal functional, being always competitive with 
the non-local HC and SM functionals. 
In particular, PGSL0.25 is performing well in case of the 
vacancy formation energies of metals, which is a severe test for most KE functionals
(here HC has a MARE of 88\% error and it also incorrectly predicts a negative 
Al-fcc vacancy formation energy; for all results see the Supporting Information). 

\begin{table}[tb]
\caption{\label{tab:siv} MARE for equilibrium volumes ($V_0$), 
bulk moduli ($B$),  
total energies ($E_0$) and density errors ($D$) for
Si phases (sc, fcc, bcc, and cd). 
Last column report the MARE for vacancy formation energies for 
fcc Al, hcp Mg, and bcc Li. 
The best results from semilocal functionals are shown in bold style, and 
a star indicates the best functional.}
%\begin{ruledtabular}
\begin{tabular}{lrrrrrrrr}
\hline\hline
          &     $V_0$      &   $B$        &     $E_0$   &  $D_0$   &  $E_{vac}$ \\
 TFW      &   9.48         & 103.24       & 2.28        & 12.9 & 133.8 \\ 
 VT84F    &   6.51         &  86.49       & 0.94        & 8.9  &  74.5 \\
 PG1      &   {\bf 2.56}         & 36.0         & 0.69        & 6.1  &  52.2 \\
 PGSL0.25 &    2.94 & {\bf 18.13}  &  {\bf 0.53} & {\bf 5.4}  & {\bf 30.5} \\ 
\hline
%-------------------------------------------------------- 
 HC      &     5.51        &   *10.41     &  0.28      & 4.4  &  88.5 \\      
 SM      &     5.11        &    19.76     & *0.21      & *2.6  &  *25.1 \\     
% SOF1 &    4.56    &   18.34   & 0.75   & 24.91 \\ 
\hline\hline  
\end{tabular}
%\end{ruledtabular}
% TFW34 &   7.3992500920088089   &  76.219821889599203   & 1.4603675979945143     & 98.630642099860594      
\end{table}

In conclusion, we have shown that it is possible to
achieve a realistic description of the KE of both metals and semiconductors
at the Laplacian semilocal level of theory, without system-dependent parameters.
This is an important result in view of 
future OF-DFT applications on large 
%NEW
and complex systems
 (e.g. hybrid interfaces). 
Moreover, semilocal functionals of the type considered here
can also easily be implemented in any 
real-space or plane-wave code.
Finally, new developments can be considered starting from the present work.
A first step concerns a further optimization of the $F(q)$ function in Eq. (\ref{pgsl}) and/or considering
more complicated functional products of both $s$ and $q$.
To this end, systems with stronger inhomogeneity (i.e. larger $q$ values) will be needed. 
A natural next step is thus the extension of the proposed functional
to interfaces or finite systems. 
%NEW
Preliminary calculations carried out on
molecular dimers have indeed shown that the PGSL0.25 functional
can describe rather well the equilibrium bond length of dimers, with an accuracy similar to HC.\cite{xia2012can}
%Test calculations carried out on
%molecular dimers have indeed shown that the PGSL0.25  functional
%describes rather well the equilibrium bond length of molecules
%(MARE=10\% to be compared with a MARE=7.5\% for HC, which 
%uses dedicated parameters for each dimer, and a special treatment 
%for the vacuum \cite{xia2012can}).
Nevertheless, for further applications in this context,
a more careful treatment of 
exponentially decaying density regions 
(where both $s$ and $q$ diverge) must be considered.

%NEW
After this work has been completed, we have acknowledged the development of the LKT GGA functional
\cite{luo2018simple}. The LKT functional form is numerically close to PG0.75 one (for $s< 1$).
An extensive assessement of various KE functionals will be presented in a
forthcoming article.

{\bf Computational Details}.
All calculations have been performed using the PROFESS 3.0 code \cite{profess}.
For a better comparison with literature results, we have chosen
the Perdew-Burke-Ernzerhof (PBE) XC functional \cite{perdewPRL96}
for simple metals \cite{xia2015single}, and the Perdew and Zunger XC LDA
parametrization \cite{perdew1981self} for semiconductors \cite{shin14}.
We use bulk-derived local pseudopotentials (BLPSs), as in Refs.
\onlinecite{xia2015single},\onlinecite{shin14}, and plane wave basis kinetic energy cutoffs of 1600 eV.
Equilibrium volumes and bulk moduli have been calculated by expanding and compressing
the optimized lattice parameters by up to about 30\% to
obtain twenty energy-volume points, fitted fitted using a birch-Murnaghan's
equation of state \cite{murnaghan1944compressibility}
expanded to 6th-order.
The reference KS density have been computed using the Abinit \cite{abinit} program. In this case denser grids
have been used (10 grid-points per angstrom) .

\textbf{Supporting Information} 

Full results for simple metals, semiconductors, Si phases, and vacancies; 
plot of the self-consistent electron densities of GaAs.

\bibliography{lap}

\end{document}